\pdfoutput=1

\documentclass[reprint,aps,prl,twocolumn,,floatfix,superscriptaddress]{revtex4-1}
\usepackage{graphicx}
\usepackage{amssymb}
\usepackage{dcolumn}
\usepackage{bm}
\usepackage{amsmath}
\usepackage{lineno}
\usepackage[english]{babel}
\usepackage[utf8]{inputenc}
\usepackage{siunitx}
\usepackage{xcolor}
\RequirePackage[T1]{fontenc}
\RequirePackage{lmodern}

\begin{document}

\title{Lightly-strained germanium quantum wells with hole mobility exceeding one million}
\author{M. Lodari}
\thanks{These authors contributed equally}
\affiliation{QuTech and Kavli Institute of Nanoscience, Delft University of Technology, PO Box 5046, 2600 GA Delft, The Netherlands}
\author{O. Kong}
\thanks{These authors contributed equally}
\affiliation{School of Physics, University of New South Wales, Sydney, New South Wales 2052, Australia}
\affiliation{ARC Centre of Excellence for Future Low-Energy Electronics Technologies, University of New South Wales, Sydney, New South Wales 2052, Australia}
\author{M. Rendell}
\affiliation{School of Physics, University of New South Wales, Sydney, New South Wales 2052, Australia}
\affiliation{ARC Centre of Excellence for Future Low-Energy Electronics Technologies, University of New South Wales, Sydney, New South Wales 2052, Australia}
\author{A. Tosato}
\affiliation{QuTech and Kavli Institute of Nanoscience, Delft University of Technology, PO Box 5046, 2600 GA Delft, The Netherlands}
\author{A. Sammak}
\affiliation{QuTech and Netherlands Organisation for Applied Scientific Research (TNO), Delft, The Netherlands.}
\author{M. Veldhorst}
\affiliation{QuTech and Kavli Institute of Nanoscience, Delft University of Technology, PO Box 5046, 2600 GA Delft, The Netherlands}
\author{A. R. Hamilton}
\affiliation{School of Physics, University of New South Wales, Sydney, New South Wales 2052, Australia}
\affiliation{ARC Centre of Excellence for Future Low-Energy Electronics Technologies, University of New South Wales, Sydney, New South Wales 2052, Australia}
\author{G. Scappucci*}
\email{g.scappucci@tudelft.nl}
\affiliation{QuTech and Kavli Institute of Nanoscience, Delft University of Technology, PO Box 5046, 2600 GA Delft, The Netherlands}

\date{\today}

\begin{abstract}
We demonstrate that a lightly-strained germanium channel ($\varepsilon_{//} = -0.41\%$) in an undoped Ge/Si$_{0.1}$Ge$_{0.9}$ heterostructure field effect transistor supports a 2D hole gas with mobility in excess of $1 \times 10^{6}$~cm$^{2}$/Vs and percolation density less than $5 \times 10^{10}$~cm$^{-2}$. This low disorder 2D hole system shows tunable fractional quantum Hall effect at low density and low magnetic field. The low-disorder and small effective mass ($0.068 m_e$) defines lightly-strained germanium as a basis to tune the strength of the spin-orbit coupling for fast and coherent quantum hardware. 
\end{abstract}

\maketitle
Quantum confined holes in germanium are emerging as a compelling  platform for quantum information processing because of several favorable properties.\cite{Scappucci_review_2021} The light hole effective mass ($\sim0.05m_e$ at zero density)\cite{lodari_light_2019} and the absence of valley degeneracy\cite{Terrazos_theory2021,Lodari_vanishingZeeman_2020} give rise to large orbital splittings in quantum dots.\cite{Hendrickx2018} The intrinsic sizable and tunable spin-orbit coupling (SOC)\cite{froning_strong_2021,bosco_squeezed_2021} enables all-electrical fast qubit driving.\cite{watzinger_germanium_2018,Hendrickx2020,froning_ultrafast_2021,wang_ultrafast_2022} Furthermore, the capability to host superconducting pairing correlations\cite{Mizokuchi_ballisticGe_2018,Hendrickx_ballisticGe_2019,Katsaros_proximityGe_2021} is promising for the co-integration of spin-qubits with superconductors in hybrid architectures for spin–spin long-distance entanglement and quantum information transfer between different qubit types.\cite{choi_spin-dependent_2000,hu_strong_2012,leijnse_hybrid_2012,leijnse_coupling_2013,hassler_exchange_2015,hoffman_universal_2016,rancic_entangling_2019}
 
Planar Ge/SiGe heterostructures are promising for scaling up to large quantum processors due to their compatibility with advanced semiconductor manufacturing.\cite{Pillarisetty2011} The low-disorder in planar Ge quantum wells\cite{Lodari_low_2021} enabled the demonstration of a four-qubit quantum processor based on hole spins in a two-by-two array of quantum dots.\cite{Hendrickx_4qubits_2021} These heterostructures featured a Si$_{0.2}$Ge$_{0.8}$ strain-relaxed buffer (SRB), resulting in quantum wells with compressive strain $\varepsilon_{//}=-0.63\%$.\cite{Sammak2019} Alternatively, higher strained Ge ($\varepsilon_{//}=-1.18~\%$) on Si$_{0.25}$Ge$_{0.75}$~SRBs enabled singlet-triplet spin qubits.\cite{Jirovec_singletriplet_2021} Lightly-strained Ge/SiGe heterostructures are unexplored and could offer potentially larger SOC because of the reduced energy splitting between heavy-holes (HH) and light holes (LH),\cite{moriya_cubic_2014} which is $\approx 17$~meV for Ge/Si$_{0.1}$Ge$_{0.9}$ compared to $\approx 51$~meV for Ge/Si$_{0.2}$Ge$_{0.8}$, respectively.\cite{Lodari_vanishingZeeman_2020,Terrazos_theory2021} As such, lightly-strained Ge is interesting for exploring faster spin-qubit driving and for topological devices. In this letter we demonstrate that lightly-strained Ge quantum wells in undoped Ge/Si$_{0.1}$Ge$_{0.9}$ support a two-dimensional hole gas (2DHG) with low disorder at low density, a prerequisite for further exploration of lightly-strained Ge quantum devices.

\begin{figure}[!ht]
\includegraphics{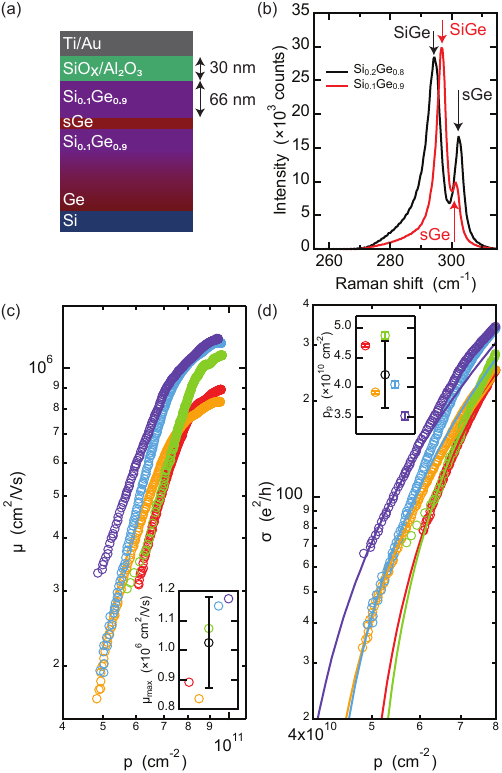}
\caption{\label{fig:1}(a) Schematic of a Ge/SiGe heterostructure field effect transistor. The strained Ge QW (sGe) is grown with the same lattice parameter to a Si$_{0.1}$Ge$_{0.9}$ SRB obtained by reverse grading, and it is separated from the native-SiO$_x$/Al$_2$O$_3$ dielectric and from the Ti/Au metallic gate stack by a 66~nm thick Si$_{0.1}$Ge$_{0.9}$ layer. (b) Intensity spectra as a function of Raman shift from a Ge/Si$_{0.2}$Ge$_{0.8}$ (black) and Ge/Si$_{0.1}$Ge$_{0.9}$ (red) heterostructures. (c) Mobility $\mu$ as a function of density $p$ at $T = 1.7$~K from five Hall bar devices from the same wafer. The inset shows the maximum mobility $\mu_{max}$ from all the devices and average value $\pm$ standard deviation (black). (d) Conductivity $\sigma_{xx}$ as a function of density $p$ (circles) and fit to percolation theory in the low density regime (solid lines). The inset shows the percolation density $p_p$ from all the devices and average value $\pm$ standard deviation (black).}
\end{figure}

\begin{figure}[!ht]
\includegraphics{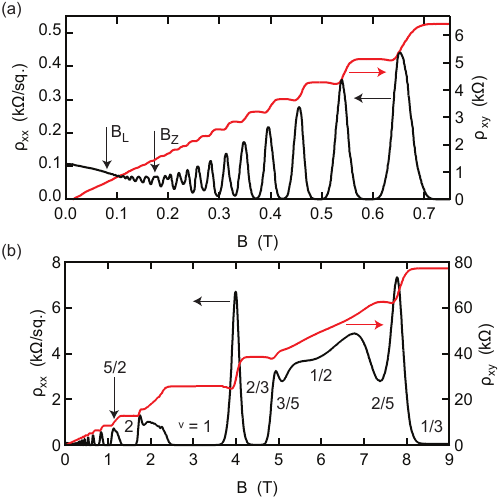}
\caption{\label{fig:2}Longitudinal resistivity $\rho_{xx}$ (black) and transverse Hall resistivity $\rho_{xy}$ (red) as a function of perpendicular magnetic field $B$ at $70$~mK, at a density of $p = 7.2 \times 10^{10}$~cm$^{-2}$ and mobility $\mu = 8.1 \times 10^{5}$~cm$^{2}$/Vs. (a) Low-field Shubnikov-de Haas oscillations from $B = 0$ to $0.75$~T, and (b) in a expanded magnetic field range from $B = 0$ to $9$~T. Onset of Landau levels ($B_L$) and Zeeman splitting ($B_Z$) are reported. Integer and fractional Landau levels labels are reported.}
\end{figure}

\begin{figure*}[!ht]
\includegraphics{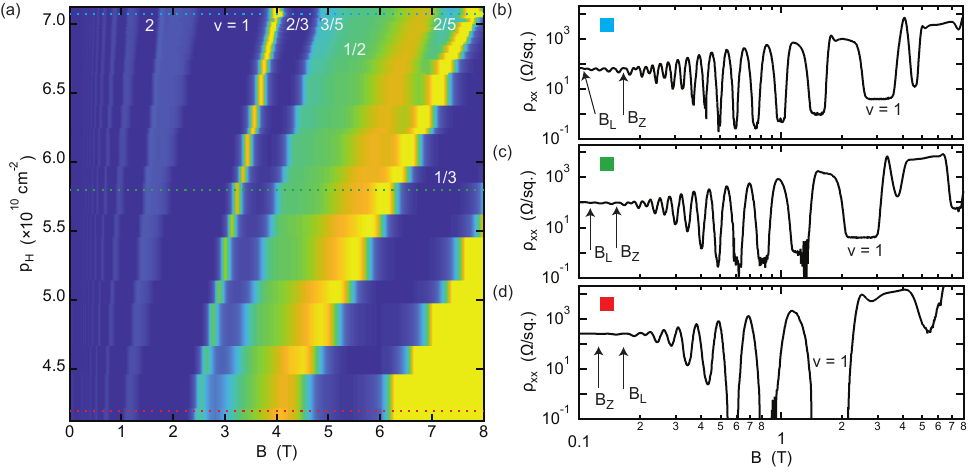}
\caption{\label{fig:3}(a) Normalized longitudinal resistivity $\rho_{xx}/\rho_{xx,0}$ as a function of magnetic field $B$ and Hall density $p$ at $T = 70$~mK. Labels of integer and fractional $\nu$ assigned from the quantum Hall effect are reported. Amplitude color scale: 0 to 60. Dashed lines correspondent to the line cuts $\rho_{xx}$ vs. $B$ at different densities are reported for (b) $p = 7.1$, (c) $5.9$, and (d) $4.2 \times 10^{10}$~cm$^{-2}$. Labels of Landau levels ($B_L$) and Zeeman splitting ($B_Z$) onset are reported.}
\end{figure*}

We grow the Ge/SiGe heterostructure by reduced-pressure chemical vapor deposition on a Si(001) wafer and then we fabricate Hall-bar shaped heterostructure field effect transistors (H-FETs) with the same process as in Refs.~\cite{Sammak2019,lodari_light_2019,Lodari_low_2021}. Here, a 16~nm strained Ge (sGe) quantum well (QW) is positioned between two strain-relaxed layers of Si$_{0.1}$Ge$_{0.9}$, at a depth of 66~nm [schematics in Fig.~\ref{fig:1}(a)].\footnote{The chemical composition of the layers is measured by secondary ion mass spectroscopy} Applying a negative DC bias to the accumulation gate $V_g$ induces a 2DHG at the Ge/Si$_{0.1}$Ge$_{0.9}$ interface. The density in the 2DHG $p$ is increased above the percolation density $p_p$ by making $V_g$ more negative. We use standard four-probe low-frequency lock-in techniques for mobility-density and magnetotransport characterization at $T=1.7$~K and $70$~mK, with excitation source-drain bias of 1~mV and 100~\si{\micro\electronvolt}, respectively. We do not measure gate to drain current leakage over the range of applied $V_g$. Figure~\ref{fig:1}(b) shows the Raman spectra measured with a 633~nm red laser to determine the strain in the Ge QW. Comparing Ge/Si$_{0.1}$Ge$_{0.9}$ (red) to a control Ge/Si$_{0.2}$Ge$_{0.8}$ (black), we observe the Raman peak from the Ge-Ge vibration mode (sGe) in the QW appearing at a lower Raman shift. Conversely, the Raman peak from the SiGe layer is appearing at a higher Raman shift. These observations are consistent with the QW in Ge/Si$_{0.1}$Ge$_{0.9}$ being less strained due to a Ge-richer SiGe SRB.\cite{Pezzoli_strain_2008} From the position of the Raman shift ($301.7$~cm$^{-1}$), we estimate a light compressive strain of $\varepsilon_{//} = -0.41\%$ for the QW in Ge/Si$_{0.1}$Ge$_{0.9}$. This is significantly lower than in Refs.~\cite{Sammak2019,Jirovec_singletriplet_2021}.

Moving on to electrical characterisation, we operate the H-FET as following. We turn on the device at $V_g \sim-0.4$~V and sweep $V_g$ to larger negative voltages ($V_g \approx -9$~V) to saturate the traps at the semiconductor/dielectric interface via charge tunneling from the quantum well, similarly to what observed in shallow Ge/Si$_{0.2}$Ge$_{0.8}$ H-FETs.\cite{Sammak2019} At these large gate voltages, the density reaches saturation ($p_{sat}$) when the Fermi level crosses the surface quantum well at the Si$_{0.1}$SiGe$_{0.9}$/dielectric interface,\cite{Su_chargetunnelling_2017} thereby screening the electric field at the sGe QW.\cite{Laroche2016} Fig.~\ref{fig:1}(c) shows the mobility $\mu$ as a function of Hall density $p$ from five H-FETs fabricated on a $2\times2$~cm$^{2}$ coupon from the center of the 100~mm wafer and measured at $T=1.7$~K. The mobility increases steeply with $p$ due to the increasing screening of scattering from remote charged impurities.\cite{Monroe1993,Gold2010,Laroche2016} At higher density ($p \geq 7\times10^{10}$~cm$^{-2}$), short range scattering from impurities within and/or in proximity of the quantum well becomes the mobility-limiting scattering mechanism.\cite{Laroche2016} We observe a maximum mobility $\mu_{max}$ in the range of $0.8-1.2 \times 10^{6}$~cm$^{2}$/Vs for $p_{sat}$ in the range of $9.43-9.64 \times 10^{10}$~cm$^{-2}$ over the five investigated H-FETs. The inset in Fig.~\ref{fig:1}(c) shows a box plot of $\mu_{max}$ across the devices, with an average value of $ (1.03 \pm 0.15)\times10^{6}$~cm$^{2}$/Vs (black), setting a benchmark for holes in buried channel transistors. Crucially, such high mobility is measured at very low density below $p = 1 \times 10^{11}$~cm$^{-2}$, a significant improvement compared to previous studies in Ge/SiGe.\cite{Sammak2019,Lodari_low_2021} 

Beyond $\mu_{max}$, $p_p$ is a key metric for characterizing the disorder potential landscape at low density, the regime relevant for quantum dot qubits. Figure~\ref{fig:1}(d) shows the conductivity $\sigma_{xx}$ (circles) as a function of density $p$ for all the investigated devices and their fit to percolation theory (lines) $\sigma \sim (p-p_p)^{1.31}$, where the exponent $1.31$ is fixed for 2D systems.\cite{Tracy2009} $p_p$ ranges from $3.5$ to $4.8 \times 10^{10}$~cm$^{-2}$. Figure~\ref{fig:1}(d) inset shows a box plot of the percolation density $p_{p}$ across the devices with an average value of $p_{p} = (4.2 \pm 0.6)\times10^{10}$~cm$^{-2}$ (black). We take these values as an upper bound for $p_p$, since we observed smaller values of $p_p$ [$(1.76 \pm 0.04) \times 10^{10}$~cm$^{-2}$ at $T=70$~mK] if the range of applied gate voltage is restricted to small voltages above the turn-on threshold.

Since Ge/S$_{0.1}$Ge$_{0.9}$ is characterized by such low level of disorder, we further explored the quantum transport proprieties of the 2DHG at $70$~mK. Figure~\ref{fig:2}(a) shows the longitudinal resistivity $\rho_{xx}$ (black) and transverse Hall resistivity $\rho_{xy}$ (red) as a function of perpendicular magnetic field $B$ up to $0.75$~T and at a Hall density of $7.2 \times 10^{10}$~cm$^{-2}$ and $\mu = 8.1 \times 10^{5}$~cm$^{2}$/Vs. We observe clear Shubnikov–de Haas (SdH) resistivity oscillations above $80$~mT. The onset for resolving the spin-degeneracy by Zeeman splitting is $0.17$~T and $\rho_{xx}$ minima reach zero already at $0.5$~T. We do not observe beatings in the SdH oscillations associated with increased Rashba spin-splitting. We speculate that such beatings are more likely to be visible at higher densities, that require the quantum well to be closer to the dielectric interface.\cite{lodari_light_2019} 

Fig.~\ref{fig:3}(b) shows the SdH oscillations at higher magnetic fields: strong minima are developed for filling factors $\nu$ with integer and fractional values. Clear plateaus are visible in $\rho_{xy}$ for $\nu = 2/3$ and $1/3$, where correspondingly $\rho_{xx}$ vanishes. Such high quality fractional quantum Hall effect (FQHE) has previously only been reported holes in modulation-doped systems at higher carrier density and, hence, at larger magnetic fields.\cite{Myronov_spinless_2015,Mironov_FQHE_2016,Mironov_FQHE_2019} Here, in undoped heterostructure, we use the top-gate to follow the evolution of FQHE states down to low density, providing avenues for studying the underlying physics. 

The color map in Fig.~\ref{fig:3}(a), measured at $T = 70$~mK, shows $\rho_{xx}$ (normalized to the value $\rho_{xx,0}$ at zero magnetic field) as a function of magnetic field $B$ and Hall density $p$ in the range of $4.1-7.1\times10^{10}$~cm$^{-2}$. Yellow and blue regions in the color map correspond to peaks and dips in the normalized $\rho_{xx}$, highlighting the density-dependent evolution of integer and fractional filling factors. All filling factors fan out towards higher magnetic field and density, and fractional filling factors are well resolved across the full investigated range of density and magnetic field.
Three line cuts from the color map are shown in Fig.~\ref{fig:3}(b--d), at decreasing density $p = 7.1$ (blue), $5.9$ (green), and $4.2\times10^{10}$~cm$^{-2}$ (red), respectively. We observe that the minima associated with fractional $\nu$ become shallower as the density is decreased, possibly because of increased level broadening by unscreened disorder and because of weaker Coulomb interactions and correlation effects.\cite{Mironov_FQHE_2016,Mironov_FQHE_2019} We also observe the distance between the onset of Shubnikov-de Haas oscillations ($B_L$) and Zeeman splitting ($B_Z$) reducing from Fig.~\ref{fig:3}(b) to Fig.~\ref{fig:3}(c). In Fig.~\ref{fig:3}(d), $B_L$ and $B_Z$ have crossed, meaning that at $p = 4.2\times10^{10}$~cm$^{-2}$ the Zeeman gap is larger than the cyclotron gap and therefore the spin susceptibility $(g^*m^*)/ m_e \geq 1$,\cite{lu_density-controlled_2017} where $m^*$ is the effective mass and $g^*$ the effective g-factor out of plane.
Indeed, from thermal activation measurement (Supplementary Material) we estimate $m^* = (0.068\pm0.001) m_e$ and $g^*= 13.95\pm0.18$ at a density of $5.8\times10^{10}$~cm$^{-2}$, corresponding to a spin susceptibility of $\approx$ 1. We note that similar values of $m^*$ and $g^*$ were reported in Ge/Si$_{0.2}$Ge$_{0.8}$\cite{lodari_light_2019,Hendrickx2018} albeit at much higher density, pointing to higher HH-LH intermixing in the lightly strained quantum wells at lower density, as expected from theory.\cite{Terrazos_theory2021}

In conclusion, we demonstrated a lightly-strained Ge/SiGe heterostrucure supporting a 2DHG with mobility in excess of one million and low percolation density (less than $5\times10^{10}$~cm$^{-2}$). Such low disorder enables measurement of FQHE at tunable low density and low magnetic fields. To mitigate the effect of traps at the interface and to suppress tunneling from the quantum well to the surface, we speculate that lightly-strained Ge channels could be positioned deeper compared to more strained channels\cite{Su_chargetunnelling_2017,Lodari_low_2021} because of the smaller band offset ($\approx 66$~meV Ge/Si$_{0.2}$Ge$_{0.9}$~vs.~$\approx 130$~meV in Ge/Si$_{0.2}$Ge$_{0.8}$). Further measurements in quantum dots, where confinement increases the HH-LH mixing, will help to elucidate the effect of reduced strain on the spin-orbit coupling in the system. The demonstration that holes can be defined in QWs with varying strain provides avenues to explore the opportunities in the germanium quantum information route.

\section{Supplemental Material}
Supplemental material is provided with measurements of the effective $g$-factor and effective mass. 

\section{Acknowledgment}
G.S. and M.V. acknowledge support through a projectruimte grant associated with the Netherlands Organization of Scientific Research (NWO). This work is partially funded by the ARC Centre of Excellence for Future Low Energy Electronics Technologies (N0. CE170100039). 

\section{Data availability}
Data sets supporting the findings of this
study are available at doi.org/10.4121/17306906.v1.

\bibliography{main.bib}

\begin{thebibliography}{38}%
\makeatletter
\providecommand \@ifxundefined [1]{%
 \@ifx{#1\undefined}
}%
\providecommand \@ifnum [1]{%
 \ifnum #1\expandafter \@firstoftwo
 \else \expandafter \@secondoftwo
 \fi
}%
\providecommand \@ifx [1]{%
 \ifx #1\expandafter \@firstoftwo
 \else \expandafter \@secondoftwo
 \fi
}%
\providecommand \natexlab [1]{#1}%
\providecommand \enquote  [1]{``#1''}%
\providecommand \bibnamefont  [1]{#1}%
\providecommand \bibfnamefont [1]{#1}%
\providecommand \citenamefont [1]{#1}%
\providecommand \href@noop [0]{\@secondoftwo}%
\providecommand \href [0]{\begingroup \@sanitize@url \@href}%
\providecommand \@href[1]{\@@startlink{#1}\@@href}%
\providecommand \@@href[1]{\endgroup#1\@@endlink}%
\providecommand \@sanitize@url [0]{\catcode `\\12\catcode `\$12\catcode
  `\&12\catcode `\#12\catcode `\^12\catcode `\_12\catcode `\%12\relax}%
\providecommand \@@startlink[1]{}%
\providecommand \@@endlink[0]{}%
\providecommand \url  [0]{\begingroup\@sanitize@url \@url }%
\providecommand \@url [1]{\endgroup\@href {#1}{\urlprefix }}%
\providecommand \urlprefix  [0]{URL }%
\providecommand \Eprint [0]{\href }%
\providecommand \doibase [0]{http://dx.doi.org/}%
\providecommand \selectlanguage [0]{\@gobble}%
\providecommand \bibinfo  [0]{\@secondoftwo}%
\providecommand \bibfield  [0]{\@secondoftwo}%
\providecommand \translation [1]{[#1]}%
\providecommand \BibitemOpen [0]{}%
\providecommand \bibitemStop [0]{}%
\providecommand \bibitemNoStop [0]{.\EOS\space}%
\providecommand \EOS [0]{\spacefactor3000\relax}%
\providecommand \BibitemShut  [1]{\csname bibitem#1\endcsname}%
\let\auto@bib@innerbib\@empty
\bibitem [{\citenamefont {Scappucci}\ \emph {et~al.}(2021)\citenamefont
  {Scappucci}, \citenamefont {Kloeffel}, \citenamefont {Zwanenburg},
  \citenamefont {Loss}, \citenamefont {Myronov}, \citenamefont {Zhang},
  \citenamefont {{De Franceschi}}, \citenamefont {Katsaros},\ and\
  \citenamefont {Veldhorst}}]{Scappucci_review_2021}%
  \BibitemOpen
  \bibfield  {author} {\bibinfo {author} {\bibfnamefont {G.}~\bibnamefont
  {Scappucci}}, \bibinfo {author} {\bibfnamefont {C.}~\bibnamefont {Kloeffel}},
  \bibinfo {author} {\bibfnamefont {F.~A.}\ \bibnamefont {Zwanenburg}},
  \bibinfo {author} {\bibfnamefont {D.}~\bibnamefont {Loss}}, \bibinfo {author}
  {\bibfnamefont {M.}~\bibnamefont {Myronov}}, \bibinfo {author} {\bibfnamefont
  {J.-J.}\ \bibnamefont {Zhang}}, \bibinfo {author} {\bibfnamefont
  {S.}~\bibnamefont {{De Franceschi}}}, \bibinfo {author} {\bibfnamefont
  {G.}~\bibnamefont {Katsaros}}, \ and\ \bibinfo {author} {\bibfnamefont
  {M.}~\bibnamefont {Veldhorst}},\ }\href {\doibase 10.1038/s41578-020-00262-z}
  {\bibfield  {journal} {\bibinfo  {journal} {Nature Reviews Materials}\
  }\textbf {\bibinfo {volume} {6}},\ \bibinfo {pages} {926} (\bibinfo {year}
  {2021})}\BibitemShut {NoStop}%
\bibitem [{\citenamefont {Lodari}\ \emph {et~al.}(2019)\citenamefont {Lodari},
  \citenamefont {Tosato}, \citenamefont {Sabbagh}, \citenamefont {Schubert},
  \citenamefont {Capellini}, \citenamefont {Sammak}, \citenamefont
  {Veldhorst},\ and\ \citenamefont {Scappucci}}]{lodari_light_2019}%
  \BibitemOpen
  \bibfield  {author} {\bibinfo {author} {\bibfnamefont {M.}~\bibnamefont
  {Lodari}}, \bibinfo {author} {\bibfnamefont {A.}~\bibnamefont {Tosato}},
  \bibinfo {author} {\bibfnamefont {D.}~\bibnamefont {Sabbagh}}, \bibinfo
  {author} {\bibfnamefont {M.~A.}\ \bibnamefont {Schubert}}, \bibinfo {author}
  {\bibfnamefont {G.}~\bibnamefont {Capellini}}, \bibinfo {author}
  {\bibfnamefont {A.}~\bibnamefont {Sammak}}, \bibinfo {author} {\bibfnamefont
  {M.}~\bibnamefont {Veldhorst}}, \ and\ \bibinfo {author} {\bibfnamefont
  {G.}~\bibnamefont {Scappucci}},\ }\href {\doibase
  10.1103/PhysRevB.100.041304} {\bibfield  {journal} {\bibinfo  {journal}
  {Physical Review B}\ }\textbf {\bibinfo {volume} {100}},\ \bibinfo {pages}
  {041304} (\bibinfo {year} {2019})}\BibitemShut {NoStop}%
\bibitem [{\citenamefont {Terrazos}\ \emph {et~al.}(2021)\citenamefont
  {Terrazos}, \citenamefont {Marcellina}, \citenamefont {Wang}, \citenamefont
  {Coppersmith}, \citenamefont {Friesen}, \citenamefont {Hamilton},
  \citenamefont {Hu}, \citenamefont {Koiller}, \citenamefont {Saraiva},
  \citenamefont {Culcer},\ and\ \citenamefont {Capaz}}]{Terrazos_theory2021}%
  \BibitemOpen
  \bibfield  {author} {\bibinfo {author} {\bibfnamefont {L.~A.}\ \bibnamefont
  {Terrazos}}, \bibinfo {author} {\bibfnamefont {E.}~\bibnamefont
  {Marcellina}}, \bibinfo {author} {\bibfnamefont {Z.}~\bibnamefont {Wang}},
  \bibinfo {author} {\bibfnamefont {S.~N.}\ \bibnamefont {Coppersmith}},
  \bibinfo {author} {\bibfnamefont {M.}~\bibnamefont {Friesen}}, \bibinfo
  {author} {\bibfnamefont {A.~R.}\ \bibnamefont {Hamilton}}, \bibinfo {author}
  {\bibfnamefont {X.}~\bibnamefont {Hu}}, \bibinfo {author} {\bibfnamefont
  {B.}~\bibnamefont {Koiller}}, \bibinfo {author} {\bibfnamefont {A.~L.}\
  \bibnamefont {Saraiva}}, \bibinfo {author} {\bibfnamefont {D.}~\bibnamefont
  {Culcer}}, \ and\ \bibinfo {author} {\bibfnamefont {R.~B.}\ \bibnamefont
  {Capaz}},\ }\href {\doibase 10.1103/PhysRevB.103.125201} {\bibfield
  {journal} {\bibinfo  {journal} {Phys. Rev. B}\ }\textbf {\bibinfo {volume}
  {103}},\ \bibinfo {pages} {125201} (\bibinfo {year} {2021})}\BibitemShut
  {NoStop}%
\bibitem [{\citenamefont {Del~Vecchio}\ \emph {et~al.}(2020)\citenamefont
  {Del~Vecchio}, \citenamefont {Lodari}, \citenamefont {Sammak}, \citenamefont
  {Scappucci},\ and\ \citenamefont
  {Moutanabbir}}]{Lodari_vanishingZeeman_2020}%
  \BibitemOpen
  \bibfield  {author} {\bibinfo {author} {\bibfnamefont {P.}~\bibnamefont
  {Del~Vecchio}}, \bibinfo {author} {\bibfnamefont {M.}~\bibnamefont {Lodari}},
  \bibinfo {author} {\bibfnamefont {A.}~\bibnamefont {Sammak}}, \bibinfo
  {author} {\bibfnamefont {G.}~\bibnamefont {Scappucci}}, \ and\ \bibinfo
  {author} {\bibfnamefont {O.}~\bibnamefont {Moutanabbir}},\ }\href {\doibase
  10.1103/PhysRevB.102.115304} {\bibfield  {journal} {\bibinfo  {journal}
  {Phys. Rev. B}\ }\textbf {\bibinfo {volume} {102}},\ \bibinfo {pages}
  {115304} (\bibinfo {year} {2020})}\BibitemShut {NoStop}%
\bibitem [{\citenamefont {Hendrickx}\ \emph {et~al.}(2018)\citenamefont
  {Hendrickx}, \citenamefont {Franke}, \citenamefont {Sammak}, \citenamefont
  {Kouwenhoven}, \citenamefont {Sabbagh}, \citenamefont {Yeoh}, \citenamefont
  {Li}, \citenamefont {Tagliaferri}, \citenamefont {Virgilio}, \citenamefont
  {Capellini}, \citenamefont {Scappucci},\ and\ \citenamefont
  {Veldhorst}}]{Hendrickx2018}%
  \BibitemOpen
  \bibfield  {author} {\bibinfo {author} {\bibfnamefont {N.~W.}\ \bibnamefont
  {Hendrickx}}, \bibinfo {author} {\bibfnamefont {D.~P.}\ \bibnamefont
  {Franke}}, \bibinfo {author} {\bibfnamefont {A.}~\bibnamefont {Sammak}},
  \bibinfo {author} {\bibfnamefont {M.}~\bibnamefont {Kouwenhoven}}, \bibinfo
  {author} {\bibfnamefont {D.}~\bibnamefont {Sabbagh}}, \bibinfo {author}
  {\bibfnamefont {L.}~\bibnamefont {Yeoh}}, \bibinfo {author} {\bibfnamefont
  {R.}~\bibnamefont {Li}}, \bibinfo {author} {\bibfnamefont {M.~L.}\
  \bibnamefont {Tagliaferri}}, \bibinfo {author} {\bibfnamefont
  {M.}~\bibnamefont {Virgilio}}, \bibinfo {author} {\bibfnamefont
  {G.}~\bibnamefont {Capellini}}, \bibinfo {author} {\bibfnamefont
  {G.}~\bibnamefont {Scappucci}}, \ and\ \bibinfo {author} {\bibfnamefont
  {M.}~\bibnamefont {Veldhorst}},\ }\href {\doibase 10.1038/s41467-018-05299-x}
  {\bibfield  {journal} {\bibinfo  {journal} {Nature Communications}\ }\textbf
  {\bibinfo {volume} {9}},\ \bibinfo {pages} {1} (\bibinfo {year} {2018})},\
  \Eprint {http://arxiv.org/abs/1801.08869} {arXiv:1801.08869} \BibitemShut
  {NoStop}%
\bibitem [{\citenamefont {Froning}\ \emph
  {et~al.}(2021{\natexlab{a}})\citenamefont {Froning}, \citenamefont
  {Rančić}, \citenamefont {Hetényi}, \citenamefont {Bosco}, \citenamefont
  {Rehmann}, \citenamefont {Li}, \citenamefont {Bakkers}, \citenamefont
  {Zwanenburg}, \citenamefont {Loss}, \citenamefont {Zumbühl},\ and\
  \citenamefont {Braakman}}]{froning_strong_2021}%
  \BibitemOpen
  \bibfield  {author} {\bibinfo {author} {\bibfnamefont {F.~N.~M.}\
  \bibnamefont {Froning}}, \bibinfo {author} {\bibfnamefont {M.~J.}\
  \bibnamefont {Rančić}}, \bibinfo {author} {\bibfnamefont {B.}~\bibnamefont
  {Hetényi}}, \bibinfo {author} {\bibfnamefont {S.}~\bibnamefont {Bosco}},
  \bibinfo {author} {\bibfnamefont {M.~K.}\ \bibnamefont {Rehmann}}, \bibinfo
  {author} {\bibfnamefont {A.}~\bibnamefont {Li}}, \bibinfo {author}
  {\bibfnamefont {E.~P. A.~M.}\ \bibnamefont {Bakkers}}, \bibinfo {author}
  {\bibfnamefont {F.~A.}\ \bibnamefont {Zwanenburg}}, \bibinfo {author}
  {\bibfnamefont {D.}~\bibnamefont {Loss}}, \bibinfo {author} {\bibfnamefont
  {D.~M.}\ \bibnamefont {Zumbühl}}, \ and\ \bibinfo {author} {\bibfnamefont
  {F.~R.}\ \bibnamefont {Braakman}},\ }\href {\doibase
  10.1103/PhysRevResearch.3.013081} {\bibfield  {journal} {\bibinfo  {journal}
  {Physical Review Research}\ }\textbf {\bibinfo {volume} {3}},\ \bibinfo
  {pages} {013081} (\bibinfo {year} {2021}{\natexlab{a}})}\BibitemShut
  {NoStop}%
\bibitem [{\citenamefont {Bosco}\ \emph {et~al.}(2021)\citenamefont {Bosco},
  \citenamefont {Benito}, \citenamefont {Adelsberger},\ and\ \citenamefont
  {Loss}}]{bosco_squeezed_2021}%
  \BibitemOpen
  \bibfield  {author} {\bibinfo {author} {\bibfnamefont {S.}~\bibnamefont
  {Bosco}}, \bibinfo {author} {\bibfnamefont {M.}~\bibnamefont {Benito}},
  \bibinfo {author} {\bibfnamefont {C.}~\bibnamefont {Adelsberger}}, \ and\
  \bibinfo {author} {\bibfnamefont {D.}~\bibnamefont {Loss}},\ }\href {\doibase
  10.1103/PhysRevB.104.115425} {\bibfield  {journal} {\bibinfo  {journal}
  {Physical Review B}\ }\textbf {\bibinfo {volume} {104}},\ \bibinfo {pages}
  {115425} (\bibinfo {year} {2021})}\BibitemShut {NoStop}%
\bibitem [{\citenamefont {Watzinger}\ \emph {et~al.}(2018)\citenamefont
  {Watzinger}, \citenamefont {Kukučka}, \citenamefont {Vukušić},
  \citenamefont {Gao}, \citenamefont {Wang}, \citenamefont {Schäffler},
  \citenamefont {Zhang},\ and\ \citenamefont
  {Katsaros}}]{watzinger_germanium_2018}%
  \BibitemOpen
  \bibfield  {author} {\bibinfo {author} {\bibfnamefont {H.}~\bibnamefont
  {Watzinger}}, \bibinfo {author} {\bibfnamefont {J.}~\bibnamefont {Kukučka}},
  \bibinfo {author} {\bibfnamefont {L.}~\bibnamefont {Vukušić}}, \bibinfo
  {author} {\bibfnamefont {F.}~\bibnamefont {Gao}}, \bibinfo {author}
  {\bibfnamefont {T.}~\bibnamefont {Wang}}, \bibinfo {author} {\bibfnamefont
  {F.}~\bibnamefont {Schäffler}}, \bibinfo {author} {\bibfnamefont {J.-J.}\
  \bibnamefont {Zhang}}, \ and\ \bibinfo {author} {\bibfnamefont
  {G.}~\bibnamefont {Katsaros}},\ }\href {\doibase 10.1038/s41467-018-06418-4}
  {\bibfield  {journal} {\bibinfo  {journal} {Nature Communications}\ }\textbf
  {\bibinfo {volume} {9}},\ \bibinfo {pages} {3902} (\bibinfo {year}
  {2018})}\BibitemShut {NoStop}%
\bibitem [{\citenamefont {Hendrickx}\ \emph {et~al.}(2020)\citenamefont
  {Hendrickx}, \citenamefont {Franke}, \citenamefont {Sammak}, \citenamefont
  {Scappucci},\ and\ \citenamefont {Veldhorst}}]{Hendrickx2020}%
  \BibitemOpen
  \bibfield  {author} {\bibinfo {author} {\bibfnamefont {N.~W.}\ \bibnamefont
  {Hendrickx}}, \bibinfo {author} {\bibfnamefont {D.~P.}\ \bibnamefont
  {Franke}}, \bibinfo {author} {\bibfnamefont {A.}~\bibnamefont {Sammak}},
  \bibinfo {author} {\bibfnamefont {G.}~\bibnamefont {Scappucci}}, \ and\
  \bibinfo {author} {\bibfnamefont {M.}~\bibnamefont {Veldhorst}},\ }\href
  {\doibase 10.1038/s41586-019-1919-3} {\bibfield  {journal} {\bibinfo
  {journal} {Nature}\ }\textbf {\bibinfo {volume} {577}},\ \bibinfo {pages}
  {487} (\bibinfo {year} {2020})},\ \Eprint {http://arxiv.org/abs/1904.11443}
  {arXiv:1904.11443} \BibitemShut {NoStop}%
\bibitem [{\citenamefont {Froning}\ \emph
  {et~al.}(2021{\natexlab{b}})\citenamefont {Froning}, \citenamefont
  {Camenzind}, \citenamefont {van~der Molen}, \citenamefont {Li}, \citenamefont
  {Bakkers}, \citenamefont {Zumbühl},\ and\ \citenamefont
  {Braakman}}]{froning_ultrafast_2021}%
  \BibitemOpen
  \bibfield  {author} {\bibinfo {author} {\bibfnamefont {F.~N.~M.}\
  \bibnamefont {Froning}}, \bibinfo {author} {\bibfnamefont {L.~C.}\
  \bibnamefont {Camenzind}}, \bibinfo {author} {\bibfnamefont {O.~A.~H.}\
  \bibnamefont {van~der Molen}}, \bibinfo {author} {\bibfnamefont
  {A.}~\bibnamefont {Li}}, \bibinfo {author} {\bibfnamefont {E.~P. A.~M.}\
  \bibnamefont {Bakkers}}, \bibinfo {author} {\bibfnamefont {D.~M.}\
  \bibnamefont {Zumbühl}}, \ and\ \bibinfo {author} {\bibfnamefont {F.~R.}\
  \bibnamefont {Braakman}},\ }\href {\doibase 10.1038/s41565-020-00828-6}
  {\bibfield  {journal} {\bibinfo  {journal} {Nature Nanotechnology}\ }\textbf
  {\bibinfo {volume} {16}},\ \bibinfo {pages} {308} (\bibinfo {year}
  {2021}{\natexlab{b}})}\BibitemShut {NoStop}%
\bibitem [{\citenamefont {Wang}\ \emph {et~al.}(2022)\citenamefont {Wang},
  \citenamefont {Xu}, \citenamefont {Gao}, \citenamefont {Liu}, \citenamefont
  {Ma}, \citenamefont {Zhang}, \citenamefont {Wang}, \citenamefont {Cao},
  \citenamefont {Wang}, \citenamefont {Zhang}, \citenamefont {Culcer},
  \citenamefont {Hu}, \citenamefont {Jiang}, \citenamefont {Li}, \citenamefont
  {Guo},\ and\ \citenamefont {Guo}}]{wang_ultrafast_2022}%
  \BibitemOpen
  \bibfield  {author} {\bibinfo {author} {\bibfnamefont {K.}~\bibnamefont
  {Wang}}, \bibinfo {author} {\bibfnamefont {G.}~\bibnamefont {Xu}}, \bibinfo
  {author} {\bibfnamefont {F.}~\bibnamefont {Gao}}, \bibinfo {author}
  {\bibfnamefont {H.}~\bibnamefont {Liu}}, \bibinfo {author} {\bibfnamefont
  {R.-L.}\ \bibnamefont {Ma}}, \bibinfo {author} {\bibfnamefont
  {X.}~\bibnamefont {Zhang}}, \bibinfo {author} {\bibfnamefont
  {Z.}~\bibnamefont {Wang}}, \bibinfo {author} {\bibfnamefont {G.}~\bibnamefont
  {Cao}}, \bibinfo {author} {\bibfnamefont {T.}~\bibnamefont {Wang}}, \bibinfo
  {author} {\bibfnamefont {J.-J.}\ \bibnamefont {Zhang}}, \bibinfo {author}
  {\bibfnamefont {D.}~\bibnamefont {Culcer}}, \bibinfo {author} {\bibfnamefont
  {X.}~\bibnamefont {Hu}}, \bibinfo {author} {\bibfnamefont {H.-W.}\
  \bibnamefont {Jiang}}, \bibinfo {author} {\bibfnamefont {H.-O.}\ \bibnamefont
  {Li}}, \bibinfo {author} {\bibfnamefont {G.-C.}\ \bibnamefont {Guo}}, \ and\
  \bibinfo {author} {\bibfnamefont {G.-P.}\ \bibnamefont {Guo}},\ }\href
  {\doibase 10.1038/s41467-021-27880-7} {\bibfield  {journal} {\bibinfo
  {journal} {Nature Communications}\ }\textbf {\bibinfo {volume} {13}},\
  \bibinfo {pages} {206} (\bibinfo {year} {2022})},\ \bibinfo {note} {arXiv:
  2006.12340}\BibitemShut {NoStop}%
\bibitem [{\citenamefont {Mizokuchi}\ \emph {et~al.}(2018)\citenamefont
  {Mizokuchi}, \citenamefont {Maurand}, \citenamefont {Vigneau}, \citenamefont
  {Myronov},\ and\ \citenamefont {{De
  Franceschi}}}]{Mizokuchi_ballisticGe_2018}%
  \BibitemOpen
  \bibfield  {author} {\bibinfo {author} {\bibfnamefont {R.}~\bibnamefont
  {Mizokuchi}}, \bibinfo {author} {\bibfnamefont {R.}~\bibnamefont {Maurand}},
  \bibinfo {author} {\bibfnamefont {F.}~\bibnamefont {Vigneau}}, \bibinfo
  {author} {\bibfnamefont {M.}~\bibnamefont {Myronov}}, \ and\ \bibinfo
  {author} {\bibfnamefont {S.}~\bibnamefont {{De Franceschi}}},\ }\href
  {\doibase 10.1021/acs.nanolett.8b01457} {\bibfield  {journal} {\bibinfo
  {journal} {Nano Letters}\ }\textbf {\bibinfo {volume} {18}},\ \bibinfo
  {pages} {4861} (\bibinfo {year} {2018})}\BibitemShut {NoStop}%
\bibitem [{\citenamefont {Hendrickx}\ \emph {et~al.}(2019)\citenamefont
  {Hendrickx}, \citenamefont {Tagliaferri}, \citenamefont {Kouwenhoven},
  \citenamefont {Li}, \citenamefont {Franke}, \citenamefont {Sammak},
  \citenamefont {Brinkman}, \citenamefont {Scappucci},\ and\ \citenamefont
  {Veldhorst}}]{Hendrickx_ballisticGe_2019}%
  \BibitemOpen
  \bibfield  {author} {\bibinfo {author} {\bibfnamefont {N.~W.}\ \bibnamefont
  {Hendrickx}}, \bibinfo {author} {\bibfnamefont {M.~L.~V.}\ \bibnamefont
  {Tagliaferri}}, \bibinfo {author} {\bibfnamefont {M.}~\bibnamefont
  {Kouwenhoven}}, \bibinfo {author} {\bibfnamefont {R.}~\bibnamefont {Li}},
  \bibinfo {author} {\bibfnamefont {D.~P.}\ \bibnamefont {Franke}}, \bibinfo
  {author} {\bibfnamefont {A.}~\bibnamefont {Sammak}}, \bibinfo {author}
  {\bibfnamefont {A.}~\bibnamefont {Brinkman}}, \bibinfo {author}
  {\bibfnamefont {G.}~\bibnamefont {Scappucci}}, \ and\ \bibinfo {author}
  {\bibfnamefont {M.}~\bibnamefont {Veldhorst}},\ }\href {\doibase
  10.1103/PhysRevB.99.075435} {\bibfield  {journal} {\bibinfo  {journal} {Phys.
  Rev. B}\ }\textbf {\bibinfo {volume} {99}},\ \bibinfo {pages} {075435}
  (\bibinfo {year} {2019})}\BibitemShut {NoStop}%
\bibitem [{\citenamefont {Aggarwal}\ \emph {et~al.}(2021)\citenamefont
  {Aggarwal}, \citenamefont {Hofmann}, \citenamefont {Jirovec}, \citenamefont
  {Prieto}, \citenamefont {Sammak}, \citenamefont {Botifoll}, \citenamefont
  {Mart\'{\i}-S\'anchez}, \citenamefont {Veldhorst}, \citenamefont {Arbiol},
  \citenamefont {Scappucci}, \citenamefont {Danon},\ and\ \citenamefont
  {Katsaros}}]{Katsaros_proximityGe_2021}%
  \BibitemOpen
  \bibfield  {author} {\bibinfo {author} {\bibfnamefont {K.}~\bibnamefont
  {Aggarwal}}, \bibinfo {author} {\bibfnamefont {A.}~\bibnamefont {Hofmann}},
  \bibinfo {author} {\bibfnamefont {D.}~\bibnamefont {Jirovec}}, \bibinfo
  {author} {\bibfnamefont {I.}~\bibnamefont {Prieto}}, \bibinfo {author}
  {\bibfnamefont {A.}~\bibnamefont {Sammak}}, \bibinfo {author} {\bibfnamefont
  {M.}~\bibnamefont {Botifoll}}, \bibinfo {author} {\bibfnamefont
  {S.}~\bibnamefont {Mart\'{\i}-S\'anchez}}, \bibinfo {author} {\bibfnamefont
  {M.}~\bibnamefont {Veldhorst}}, \bibinfo {author} {\bibfnamefont
  {J.}~\bibnamefont {Arbiol}}, \bibinfo {author} {\bibfnamefont
  {G.}~\bibnamefont {Scappucci}}, \bibinfo {author} {\bibfnamefont
  {J.}~\bibnamefont {Danon}}, \ and\ \bibinfo {author} {\bibfnamefont
  {G.}~\bibnamefont {Katsaros}},\ }\href {\doibase
  10.1103/PhysRevResearch.3.L022005} {\bibfield  {journal} {\bibinfo  {journal}
  {Phys. Rev. Research}\ }\textbf {\bibinfo {volume} {3}},\ \bibinfo {pages}
  {L022005} (\bibinfo {year} {2021})}\BibitemShut {NoStop}%
\bibitem [{\citenamefont {Choi}\ \emph {et~al.}(2000)\citenamefont {Choi},
  \citenamefont {Bruder},\ and\ \citenamefont
  {Loss}}]{choi_spin-dependent_2000}%
  \BibitemOpen
  \bibfield  {author} {\bibinfo {author} {\bibfnamefont {M.-S.}\ \bibnamefont
  {Choi}}, \bibinfo {author} {\bibfnamefont {C.}~\bibnamefont {Bruder}}, \ and\
  \bibinfo {author} {\bibfnamefont {D.}~\bibnamefont {Loss}},\ }\href {\doibase
  10.1103/PhysRevB.62.13569} {\bibfield  {journal} {\bibinfo  {journal}
  {Physical Review B}\ }\textbf {\bibinfo {volume} {62}},\ \bibinfo {pages}
  {13569} (\bibinfo {year} {2000})}\BibitemShut {NoStop}%
\bibitem [{\citenamefont {Hu}\ \emph {et~al.}(2012)\citenamefont {Hu},
  \citenamefont {Liu},\ and\ \citenamefont {Nori}}]{hu_strong_2012}%
  \BibitemOpen
  \bibfield  {author} {\bibinfo {author} {\bibfnamefont {X.}~\bibnamefont
  {Hu}}, \bibinfo {author} {\bibfnamefont {Y.-x.}\ \bibnamefont {Liu}}, \ and\
  \bibinfo {author} {\bibfnamefont {F.}~\bibnamefont {Nori}},\ }\href {\doibase
  10.1103/PhysRevB.86.035314} {\bibfield  {journal} {\bibinfo  {journal}
  {Physical Review B}\ }\textbf {\bibinfo {volume} {86}},\ \bibinfo {pages}
  {035314} (\bibinfo {year} {2012})}\BibitemShut {NoStop}%
\bibitem [{\citenamefont {Leijnse}\ and\ \citenamefont
  {Flensberg}(2012)}]{leijnse_hybrid_2012}%
  \BibitemOpen
  \bibfield  {author} {\bibinfo {author} {\bibfnamefont {M.}~\bibnamefont
  {Leijnse}}\ and\ \bibinfo {author} {\bibfnamefont {K.}~\bibnamefont
  {Flensberg}},\ }\href {\doibase 10.1103/PhysRevB.86.104511} {\bibfield
  {journal} {\bibinfo  {journal} {Physical Review B}\ }\textbf {\bibinfo
  {volume} {86}},\ \bibinfo {pages} {104511} (\bibinfo {year}
  {2012})}\BibitemShut {NoStop}%
\bibitem [{\citenamefont {Leijnse}\ and\ \citenamefont
  {Flensberg}(2013)}]{leijnse_coupling_2013}%
  \BibitemOpen
  \bibfield  {author} {\bibinfo {author} {\bibfnamefont {M.}~\bibnamefont
  {Leijnse}}\ and\ \bibinfo {author} {\bibfnamefont {K.}~\bibnamefont
  {Flensberg}},\ }\href {\doibase 10.1103/PhysRevLett.111.060501} {\bibfield
  {journal} {\bibinfo  {journal} {Physical Review Letters}\ }\textbf {\bibinfo
  {volume} {111}},\ \bibinfo {pages} {060501} (\bibinfo {year}
  {2013})}\BibitemShut {NoStop}%
\bibitem [{\citenamefont {Hassler}\ \emph {et~al.}(2015)\citenamefont
  {Hassler}, \citenamefont {Catelani},\ and\ \citenamefont
  {Bluhm}}]{hassler_exchange_2015}%
  \BibitemOpen
  \bibfield  {author} {\bibinfo {author} {\bibfnamefont {F.}~\bibnamefont
  {Hassler}}, \bibinfo {author} {\bibfnamefont {G.}~\bibnamefont {Catelani}}, \
  and\ \bibinfo {author} {\bibfnamefont {H.}~\bibnamefont {Bluhm}},\ }\href
  {\doibase 10.1103/PhysRevB.92.235401} {\bibfield  {journal} {\bibinfo
  {journal} {Physical Review B}\ }\textbf {\bibinfo {volume} {92}},\ \bibinfo
  {pages} {235401} (\bibinfo {year} {2015})}\BibitemShut {NoStop}%
\bibitem [{\citenamefont {Hoffman}\ \emph {et~al.}(2016)\citenamefont
  {Hoffman}, \citenamefont {Schrade}, \citenamefont {Klinovaja},\ and\
  \citenamefont {Loss}}]{hoffman_universal_2016}%
  \BibitemOpen
  \bibfield  {author} {\bibinfo {author} {\bibfnamefont {S.}~\bibnamefont
  {Hoffman}}, \bibinfo {author} {\bibfnamefont {C.}~\bibnamefont {Schrade}},
  \bibinfo {author} {\bibfnamefont {J.}~\bibnamefont {Klinovaja}}, \ and\
  \bibinfo {author} {\bibfnamefont {D.}~\bibnamefont {Loss}},\ }\href {\doibase
  10.1103/PhysRevB.94.045316} {\bibfield  {journal} {\bibinfo  {journal}
  {Physical Review B}\ }\textbf {\bibinfo {volume} {94}},\ \bibinfo {pages}
  {045316} (\bibinfo {year} {2016})}\BibitemShut {NoStop}%
\bibitem [{\citenamefont {Rančić}\ \emph {et~al.}(2019)\citenamefont
  {Rančić}, \citenamefont {Hoffman}, \citenamefont {Schrade}, \citenamefont
  {Klinovaja},\ and\ \citenamefont {Loss}}]{rancic_entangling_2019}%
  \BibitemOpen
  \bibfield  {author} {\bibinfo {author} {\bibfnamefont {M.~J.}\ \bibnamefont
  {Rančić}}, \bibinfo {author} {\bibfnamefont {S.}~\bibnamefont {Hoffman}},
  \bibinfo {author} {\bibfnamefont {C.}~\bibnamefont {Schrade}}, \bibinfo
  {author} {\bibfnamefont {J.}~\bibnamefont {Klinovaja}}, \ and\ \bibinfo
  {author} {\bibfnamefont {D.}~\bibnamefont {Loss}},\ }\href {\doibase
  10.1103/PhysRevB.99.165306} {\bibfield  {journal} {\bibinfo  {journal}
  {Physical Review B}\ }\textbf {\bibinfo {volume} {99}},\ \bibinfo {pages}
  {165306} (\bibinfo {year} {2019})}\BibitemShut {NoStop}%
\bibitem [{\citenamefont {Pillarisetty}(2011)}]{Pillarisetty2011}%
  \BibitemOpen
  \bibfield  {author} {\bibinfo {author} {\bibfnamefont {R.}~\bibnamefont
  {Pillarisetty}},\ }\href {\doibase 10.1038/nature10678} {\bibfield  {journal}
  {\bibinfo  {journal} {Nature}\ }\textbf {\bibinfo {volume} {479}},\ \bibinfo
  {pages} {324} (\bibinfo {year} {2011})}\BibitemShut {NoStop}%
\bibitem [{\citenamefont {Lodari}\ \emph {et~al.}(2021)\citenamefont {Lodari},
  \citenamefont {Hendrickx}, \citenamefont {Lawrie}, \citenamefont {Hsiao},
  \citenamefont {Vandersypen}, \citenamefont {Sammak}, \citenamefont
  {Veldhorst},\ and\ \citenamefont {Scappucci}}]{Lodari_low_2021}%
  \BibitemOpen
  \bibfield  {author} {\bibinfo {author} {\bibfnamefont {M.}~\bibnamefont
  {Lodari}}, \bibinfo {author} {\bibfnamefont {N.~W.}\ \bibnamefont
  {Hendrickx}}, \bibinfo {author} {\bibfnamefont {W.~I.~L.}\ \bibnamefont
  {Lawrie}}, \bibinfo {author} {\bibfnamefont {T.-K.}\ \bibnamefont {Hsiao}},
  \bibinfo {author} {\bibfnamefont {L.~M.~K.}\ \bibnamefont {Vandersypen}},
  \bibinfo {author} {\bibfnamefont {A.}~\bibnamefont {Sammak}}, \bibinfo
  {author} {\bibfnamefont {M.}~\bibnamefont {Veldhorst}}, \ and\ \bibinfo
  {author} {\bibfnamefont {G.}~\bibnamefont {Scappucci}},\ }\href {\doibase
  10.1088/2633-4356/abcd82} {\bibfield  {journal} {\bibinfo  {journal}
  {Materials for Quantum Technology}\ }\textbf {\bibinfo {volume} {1}},\
  \bibinfo {pages} {011002} (\bibinfo {year} {2021})},\ \Eprint
  {http://arxiv.org/abs/2007.06328} {arXiv:2007.06328} \BibitemShut {NoStop}%
\bibitem [{\citenamefont {Hendrickx}\ \emph {et~al.}(2021)\citenamefont
  {Hendrickx}, \citenamefont {Lawrie}, \citenamefont {Russ}, \citenamefont {van
  Riggelen}, \citenamefont {de~Snoo}, \citenamefont {Schouten}, \citenamefont
  {Sammak}, \citenamefont {Scappucci},\ and\ \citenamefont
  {Veldhorst}}]{Hendrickx_4qubits_2021}%
  \BibitemOpen
  \bibfield  {author} {\bibinfo {author} {\bibfnamefont {N.~W.}\ \bibnamefont
  {Hendrickx}}, \bibinfo {author} {\bibfnamefont {W.~I.~L.}\ \bibnamefont
  {Lawrie}}, \bibinfo {author} {\bibfnamefont {M.}~\bibnamefont {Russ}},
  \bibinfo {author} {\bibfnamefont {F.}~\bibnamefont {van Riggelen}}, \bibinfo
  {author} {\bibfnamefont {S.~L.}\ \bibnamefont {de~Snoo}}, \bibinfo {author}
  {\bibfnamefont {R.~N.}\ \bibnamefont {Schouten}}, \bibinfo {author}
  {\bibfnamefont {A.}~\bibnamefont {Sammak}}, \bibinfo {author} {\bibfnamefont
  {G.}~\bibnamefont {Scappucci}}, \ and\ \bibinfo {author} {\bibfnamefont
  {M.}~\bibnamefont {Veldhorst}},\ }\href {\doibase 10.1038/s41586-021-03332-6}
  {\bibfield  {journal} {\bibinfo  {journal} {Nature}\ }\textbf {\bibinfo
  {volume} {591}},\ \bibinfo {pages} {580} (\bibinfo {year}
  {2021})}\BibitemShut {NoStop}%
\bibitem [{\citenamefont {Sammak}\ \emph {et~al.}(2019)\citenamefont {Sammak},
  \citenamefont {Sabbagh}, \citenamefont {Hendrickx}, \citenamefont {Lodari},
  \citenamefont {{Paquelet Wuetz}}, \citenamefont {Tosato}, \citenamefont
  {Yeoh}, \citenamefont {Bollani}, \citenamefont {Virgilio}, \citenamefont
  {Schubert}, \citenamefont {Zaumseil}, \citenamefont {Capellini},
  \citenamefont {Veldhorst},\ and\ \citenamefont {Scappucci}}]{Sammak2019}%
  \BibitemOpen
  \bibfield  {author} {\bibinfo {author} {\bibfnamefont {A.}~\bibnamefont
  {Sammak}}, \bibinfo {author} {\bibfnamefont {D.}~\bibnamefont {Sabbagh}},
  \bibinfo {author} {\bibfnamefont {N.~W.}\ \bibnamefont {Hendrickx}}, \bibinfo
  {author} {\bibfnamefont {M.}~\bibnamefont {Lodari}}, \bibinfo {author}
  {\bibfnamefont {B.}~\bibnamefont {{Paquelet Wuetz}}}, \bibinfo {author}
  {\bibfnamefont {A.}~\bibnamefont {Tosato}}, \bibinfo {author} {\bibfnamefont
  {L.~R.}\ \bibnamefont {Yeoh}}, \bibinfo {author} {\bibfnamefont
  {M.}~\bibnamefont {Bollani}}, \bibinfo {author} {\bibfnamefont
  {M.}~\bibnamefont {Virgilio}}, \bibinfo {author} {\bibfnamefont {M.~A.}\
  \bibnamefont {Schubert}}, \bibinfo {author} {\bibfnamefont {P.}~\bibnamefont
  {Zaumseil}}, \bibinfo {author} {\bibfnamefont {G.}~\bibnamefont {Capellini}},
  \bibinfo {author} {\bibfnamefont {M.}~\bibnamefont {Veldhorst}}, \ and\
  \bibinfo {author} {\bibfnamefont {G.}~\bibnamefont {Scappucci}},\ }\href
  {\doibase 10.1002/adfm.201807613} {\bibfield  {journal} {\bibinfo  {journal}
  {Advanced Functional Materials}\ }\textbf {\bibinfo {volume} {29}},\ \bibinfo
  {pages} {1} (\bibinfo {year} {2019})}\BibitemShut {NoStop}%
\bibitem [{\citenamefont {Jirovec}\ \emph {et~al.}(2021)\citenamefont
  {Jirovec}, \citenamefont {Hofmann}, \citenamefont {Ballabio}, \citenamefont
  {Mutter}, \citenamefont {Tavani}, \citenamefont {Botifoll}, \citenamefont
  {Crippa}, \citenamefont {Kukucka}, \citenamefont {Sagi}, \citenamefont
  {Martins}, \citenamefont {Saez-Mollejo}, \citenamefont {Prieto},
  \citenamefont {Borovkov}, \citenamefont {Arbiol}, \citenamefont {Chrastina},
  \citenamefont {Isella},\ and\ \citenamefont
  {Katsaros}}]{Jirovec_singletriplet_2021}%
  \BibitemOpen
  \bibfield  {author} {\bibinfo {author} {\bibfnamefont {D.}~\bibnamefont
  {Jirovec}}, \bibinfo {author} {\bibfnamefont {A.}~\bibnamefont {Hofmann}},
  \bibinfo {author} {\bibfnamefont {A.}~\bibnamefont {Ballabio}}, \bibinfo
  {author} {\bibfnamefont {P.~M.}\ \bibnamefont {Mutter}}, \bibinfo {author}
  {\bibfnamefont {G.}~\bibnamefont {Tavani}}, \bibinfo {author} {\bibfnamefont
  {M.}~\bibnamefont {Botifoll}}, \bibinfo {author} {\bibfnamefont
  {A.}~\bibnamefont {Crippa}}, \bibinfo {author} {\bibfnamefont
  {J.}~\bibnamefont {Kukucka}}, \bibinfo {author} {\bibfnamefont
  {O.}~\bibnamefont {Sagi}}, \bibinfo {author} {\bibfnamefont {F.}~\bibnamefont
  {Martins}}, \bibinfo {author} {\bibfnamefont {J.}~\bibnamefont
  {Saez-Mollejo}}, \bibinfo {author} {\bibfnamefont {I.}~\bibnamefont
  {Prieto}}, \bibinfo {author} {\bibfnamefont {M.}~\bibnamefont {Borovkov}},
  \bibinfo {author} {\bibfnamefont {J.}~\bibnamefont {Arbiol}}, \bibinfo
  {author} {\bibfnamefont {D.}~\bibnamefont {Chrastina}}, \bibinfo {author}
  {\bibfnamefont {G.}~\bibnamefont {Isella}}, \ and\ \bibinfo {author}
  {\bibfnamefont {G.}~\bibnamefont {Katsaros}},\ }\href {\doibase
  10.1038/s41563-021-01022-2} {\bibfield  {journal} {\bibinfo  {journal}
  {Nature Materials}\ }\textbf {\bibinfo {volume} {20}},\ \bibinfo {pages}
  {1106} (\bibinfo {year} {2021})}\BibitemShut {NoStop}%
\bibitem [{\citenamefont {Moriya}\ \emph {et~al.}(2014)\citenamefont {Moriya},
  \citenamefont {Sawano}, \citenamefont {Hoshi}, \citenamefont {Masubuchi},
  \citenamefont {Shiraki}, \citenamefont {Wild}, \citenamefont {Neumann},
  \citenamefont {Abstreiter}, \citenamefont {Bougeard}, \citenamefont {Koga},\
  and\ \citenamefont {Machida}}]{moriya_cubic_2014}%
  \BibitemOpen
  \bibfield  {author} {\bibinfo {author} {\bibfnamefont {R.}~\bibnamefont
  {Moriya}}, \bibinfo {author} {\bibfnamefont {K.}~\bibnamefont {Sawano}},
  \bibinfo {author} {\bibfnamefont {Y.}~\bibnamefont {Hoshi}}, \bibinfo
  {author} {\bibfnamefont {S.}~\bibnamefont {Masubuchi}}, \bibinfo {author}
  {\bibfnamefont {Y.}~\bibnamefont {Shiraki}}, \bibinfo {author} {\bibfnamefont
  {A.}~\bibnamefont {Wild}}, \bibinfo {author} {\bibfnamefont {C.}~\bibnamefont
  {Neumann}}, \bibinfo {author} {\bibfnamefont {G.}~\bibnamefont {Abstreiter}},
  \bibinfo {author} {\bibfnamefont {D.}~\bibnamefont {Bougeard}}, \bibinfo
  {author} {\bibfnamefont {T.}~\bibnamefont {Koga}}, \ and\ \bibinfo {author}
  {\bibfnamefont {T.}~\bibnamefont {Machida}},\ }\href {\doibase
  10.1103/PhysRevLett.113.086601} {\bibfield  {journal} {\bibinfo  {journal}
  {Physical Review Letters}\ }\textbf {\bibinfo {volume} {113}},\ \bibinfo
  {pages} {086601} (\bibinfo {year} {2014})}\BibitemShut {NoStop}%
\bibitem [{Note1()}]{Note1}%
  \BibitemOpen
  \bibinfo {note} {The chemical composition of the layers is measured by
  secondary ion mass spectroscopy}\BibitemShut {NoStop}%
\bibitem [{\citenamefont {Pezzoli}\ \emph {et~al.}(2008)\citenamefont
  {Pezzoli}, \citenamefont {Bonera}, \citenamefont {Grilli}, \citenamefont
  {Guzzi}, \citenamefont {Sanguinetti}, \citenamefont {Chrastina},
  \citenamefont {Isella}, \citenamefont {{Von K{\"{a}}nel}}, \citenamefont
  {Wintersberger}, \citenamefont {Stangl},\ and\ \citenamefont
  {Bauer}}]{Pezzoli_strain_2008}%
  \BibitemOpen
  \bibfield  {author} {\bibinfo {author} {\bibfnamefont {F.}~\bibnamefont
  {Pezzoli}}, \bibinfo {author} {\bibfnamefont {E.}~\bibnamefont {Bonera}},
  \bibinfo {author} {\bibfnamefont {E.}~\bibnamefont {Grilli}}, \bibinfo
  {author} {\bibfnamefont {M.}~\bibnamefont {Guzzi}}, \bibinfo {author}
  {\bibfnamefont {S.}~\bibnamefont {Sanguinetti}}, \bibinfo {author}
  {\bibfnamefont {D.}~\bibnamefont {Chrastina}}, \bibinfo {author}
  {\bibfnamefont {G.}~\bibnamefont {Isella}}, \bibinfo {author} {\bibfnamefont
  {H.}~\bibnamefont {{Von K{\"{a}}nel}}}, \bibinfo {author} {\bibfnamefont
  {E.}~\bibnamefont {Wintersberger}}, \bibinfo {author} {\bibfnamefont
  {J.}~\bibnamefont {Stangl}}, \ and\ \bibinfo {author} {\bibfnamefont
  {G.}~\bibnamefont {Bauer}},\ }\href {\doibase 10.1063/1.2913052} {\bibfield
  {journal} {\bibinfo  {journal} {Journal of Applied Physics}\ }\textbf
  {\bibinfo {volume} {103}} (\bibinfo {year} {2008}),\
  10.1063/1.2913052}\BibitemShut {NoStop}%
\bibitem [{\citenamefont {Su}\ \emph {et~al.}(2017)\citenamefont {Su},
  \citenamefont {Chuang}, \citenamefont {Liu}, \citenamefont {Li},\ and\
  \citenamefont {Lu}}]{Su_chargetunnelling_2017}%
  \BibitemOpen
  \bibfield  {author} {\bibinfo {author} {\bibfnamefont {Y.-H.}\ \bibnamefont
  {Su}}, \bibinfo {author} {\bibfnamefont {Y.}~\bibnamefont {Chuang}}, \bibinfo
  {author} {\bibfnamefont {C.-Y.}\ \bibnamefont {Liu}}, \bibinfo {author}
  {\bibfnamefont {J.-Y.}\ \bibnamefont {Li}}, \ and\ \bibinfo {author}
  {\bibfnamefont {T.-M.}\ \bibnamefont {Lu}},\ }\href {\doibase
  10.1103/PhysRevMaterials.1.044601} {\bibfield  {journal} {\bibinfo  {journal}
  {Phys. Rev. Materials}\ }\textbf {\bibinfo {volume} {1}},\ \bibinfo {pages}
  {044601} (\bibinfo {year} {2017})}\BibitemShut {NoStop}%
\bibitem [{\citenamefont {Laroche}\ \emph {et~al.}(2016)\citenamefont
  {Laroche}, \citenamefont {Huang}, \citenamefont {Chuang}, \citenamefont {Li},
  \citenamefont {Liu},\ and\ \citenamefont {Lu}}]{Laroche2016}%
  \BibitemOpen
  \bibfield  {author} {\bibinfo {author} {\bibfnamefont {D.}~\bibnamefont
  {Laroche}}, \bibinfo {author} {\bibfnamefont {S.~H.}\ \bibnamefont {Huang}},
  \bibinfo {author} {\bibfnamefont {Y.}~\bibnamefont {Chuang}}, \bibinfo
  {author} {\bibfnamefont {J.~Y.}\ \bibnamefont {Li}}, \bibinfo {author}
  {\bibfnamefont {C.~W.}\ \bibnamefont {Liu}}, \ and\ \bibinfo {author}
  {\bibfnamefont {T.~M.}\ \bibnamefont {Lu}},\ }\href {\doibase
  10.1063/1.4953399} {\bibfield  {journal} {\bibinfo  {journal} {Applied
  Physics Letters}\ }\textbf {\bibinfo {volume} {108}} (\bibinfo {year}
  {2016}),\ 10.1063/1.4953399}\BibitemShut {NoStop}%
\bibitem [{\citenamefont {Monroe}(1993)}]{Monroe1993}%
  \BibitemOpen
  \bibfield  {author} {\bibinfo {author} {\bibfnamefont {D.}~\bibnamefont
  {Monroe}},\ }\href {\doibase 10.1116/1.586471} {\bibfield  {journal}
  {\bibinfo  {journal} {Journal of Vacuum Science $\&$ Technology B:
  Microelectronics and Nanometer Structures}\ }\textbf {\bibinfo {volume}
  {11}},\ \bibinfo {pages} {1731} (\bibinfo {year} {1993})}\BibitemShut
  {NoStop}%
\bibitem [{\citenamefont {Gold}(2010)}]{Gold2010}%
  \BibitemOpen
  \bibfield  {author} {\bibinfo {author} {\bibfnamefont {A.}~\bibnamefont
  {Gold}},\ }\href {\doibase 10.1063/1.3482058} {\bibfield  {journal} {\bibinfo
   {journal} {Journal of Applied Physics}\ }\textbf {\bibinfo {volume} {108}}
  (\bibinfo {year} {2010}),\ 10.1063/1.3482058}\BibitemShut {NoStop}%
\bibitem [{\citenamefont {Tracy}\ \emph {et~al.}(2009)\citenamefont {Tracy},
  \citenamefont {Hwang}, \citenamefont {Eng}, \citenamefont {{Ten Eyck}},
  \citenamefont {Nordberg}, \citenamefont {Childs}, \citenamefont {Carroll},
  \citenamefont {Lilly},\ and\ \citenamefont {{Das Sarma}}}]{Tracy2009}%
  \BibitemOpen
  \bibfield  {author} {\bibinfo {author} {\bibfnamefont {L.~A.}\ \bibnamefont
  {Tracy}}, \bibinfo {author} {\bibfnamefont {E.~H.}\ \bibnamefont {Hwang}},
  \bibinfo {author} {\bibfnamefont {K.}~\bibnamefont {Eng}}, \bibinfo {author}
  {\bibfnamefont {G.~A.}\ \bibnamefont {{Ten Eyck}}}, \bibinfo {author}
  {\bibfnamefont {E.~P.}\ \bibnamefont {Nordberg}}, \bibinfo {author}
  {\bibfnamefont {K.}~\bibnamefont {Childs}}, \bibinfo {author} {\bibfnamefont
  {M.~S.}\ \bibnamefont {Carroll}}, \bibinfo {author} {\bibfnamefont {M.~P.}\
  \bibnamefont {Lilly}}, \ and\ \bibinfo {author} {\bibfnamefont
  {S.}~\bibnamefont {{Das Sarma}}},\ }\href {\doibase
  10.1103/PhysRevB.79.235307} {\bibfield  {journal} {\bibinfo  {journal}
  {Physical Review B - Condensed Matter and Materials Physics}\ }\textbf
  {\bibinfo {volume} {79}},\ \bibinfo {pages} {1} (\bibinfo {year}
  {2009})}\BibitemShut {NoStop}%
\bibitem [{\citenamefont {Shi}\ \emph {et~al.}(2015)\citenamefont {Shi},
  \citenamefont {Zudov}, \citenamefont {Morrison},\ and\ \citenamefont
  {Myronov}}]{Myronov_spinless_2015}%
  \BibitemOpen
  \bibfield  {author} {\bibinfo {author} {\bibfnamefont {Q.}~\bibnamefont
  {Shi}}, \bibinfo {author} {\bibfnamefont {M.~A.}\ \bibnamefont {Zudov}},
  \bibinfo {author} {\bibfnamefont {C.}~\bibnamefont {Morrison}}, \ and\
  \bibinfo {author} {\bibfnamefont {M.}~\bibnamefont {Myronov}},\ }\href
  {\doibase 10.1103/PhysRevB.91.241303} {\bibfield  {journal} {\bibinfo
  {journal} {Phys. Rev. B}\ }\textbf {\bibinfo {volume} {91}},\ \bibinfo
  {pages} {241303} (\bibinfo {year} {2015})}\BibitemShut {NoStop}%
\bibitem [{\citenamefont {Mironov}\ \emph {et~al.}(2016)\citenamefont
  {Mironov}, \citenamefont {d'Ambrumenil}, \citenamefont {Dobbie},
  \citenamefont {Leadley}, \citenamefont {Suslov},\ and\ \citenamefont
  {Green}}]{Mironov_FQHE_2016}%
  \BibitemOpen
  \bibfield  {author} {\bibinfo {author} {\bibfnamefont {O.~A.}\ \bibnamefont
  {Mironov}}, \bibinfo {author} {\bibfnamefont {N.}~\bibnamefont
  {d'Ambrumenil}}, \bibinfo {author} {\bibfnamefont {A.}~\bibnamefont
  {Dobbie}}, \bibinfo {author} {\bibfnamefont {D.~R.}\ \bibnamefont {Leadley}},
  \bibinfo {author} {\bibfnamefont {A.~V.}\ \bibnamefont {Suslov}}, \ and\
  \bibinfo {author} {\bibfnamefont {E.}~\bibnamefont {Green}},\ }\href
  {\doibase 10.1103/PhysRevLett.116.176802} {\bibfield  {journal} {\bibinfo
  {journal} {Phys. Rev. Lett.}\ }\textbf {\bibinfo {volume} {116}},\ \bibinfo
  {pages} {176802} (\bibinfo {year} {2016})}\BibitemShut {NoStop}%
\bibitem [{\citenamefont {Berkutov}\ \emph {et~al.}(2019)\citenamefont
  {Berkutov}, \citenamefont {Andrievskii}, \citenamefont {Kolesnichenko},\ and\
  \citenamefont {Mironov}}]{Mironov_FQHE_2019}%
  \BibitemOpen
  \bibfield  {author} {\bibinfo {author} {\bibfnamefont {I.~B.}\ \bibnamefont
  {Berkutov}}, \bibinfo {author} {\bibfnamefont {V.~V.}\ \bibnamefont
  {Andrievskii}}, \bibinfo {author} {\bibfnamefont {Y.~A.}\ \bibnamefont
  {Kolesnichenko}}, \ and\ \bibinfo {author} {\bibfnamefont {O.~A.}\
  \bibnamefont {Mironov}},\ }\href {\doibase 10.1063/10.0000126} {\bibfield
  {journal} {\bibinfo  {journal} {Low Temperature Physics}\ }\textbf {\bibinfo
  {volume} {45}},\ \bibinfo {pages} {1202} (\bibinfo {year}
  {2019})}\BibitemShut {NoStop}%
\bibitem [{\citenamefont {Lu}\ \emph {et~al.}(2017)\citenamefont {Lu},
  \citenamefont {Tracy}, \citenamefont {Laroche}, \citenamefont {Huang},
  \citenamefont {Chuang}, \citenamefont {Su}, \citenamefont {Li},\ and\
  \citenamefont {Liu}}]{lu_density-controlled_2017}%
  \BibitemOpen
  \bibfield  {author} {\bibinfo {author} {\bibfnamefont {T.~M.}\ \bibnamefont
  {Lu}}, \bibinfo {author} {\bibfnamefont {L.~A.}\ \bibnamefont {Tracy}},
  \bibinfo {author} {\bibfnamefont {D.}~\bibnamefont {Laroche}}, \bibinfo
  {author} {\bibfnamefont {S.-H.}\ \bibnamefont {Huang}}, \bibinfo {author}
  {\bibfnamefont {Y.}~\bibnamefont {Chuang}}, \bibinfo {author} {\bibfnamefont
  {Y.-H.}\ \bibnamefont {Su}}, \bibinfo {author} {\bibfnamefont {J.-Y.}\
  \bibnamefont {Li}}, \ and\ \bibinfo {author} {\bibfnamefont {C.~W.}\
  \bibnamefont {Liu}},\ }\href {\doibase 10.1038/s41598-017-02757-2} {\bibfield
   {journal} {\bibinfo  {journal} {Scientific Reports}\ }\textbf {\bibinfo
  {volume} {7}},\ \bibinfo {pages} {2468} (\bibinfo {year} {2017})}\BibitemShut
  {NoStop}%
\end{thebibliography}%


\begin{thebibliography}{0}%
\makeatletter
\providecommand \@ifxundefined [1]{%
 \@ifx{#1\undefined}
}%
\providecommand \@ifnum [1]{%
 \ifnum #1\expandafter \@firstoftwo
 \else \expandafter \@secondoftwo
 \fi
}%
\providecommand \@ifx [1]{%
 \ifx #1\expandafter \@firstoftwo
 \else \expandafter \@secondoftwo
 \fi
}%
\providecommand \natexlab [1]{#1}%
\providecommand \enquote  [1]{``#1''}%
\providecommand \bibnamefont  [1]{#1}%
\providecommand \bibfnamefont [1]{#1}%
\providecommand \citenamefont [1]{#1}%
\providecommand \href@noop [0]{\@secondoftwo}%
\providecommand \href [0]{\begingroup \@sanitize@url \@href}%
\providecommand \@href[1]{\@@startlink{#1}\@@href}%
\providecommand \@@href[1]{\endgroup#1\@@endlink}%
\providecommand \@sanitize@url [0]{\catcode `\\12\catcode `\$12\catcode
  `\&12\catcode `\#12\catcode `\^12\catcode `\_12\catcode `\%12\relax}%
\providecommand \@@startlink[1]{}%
\providecommand \@@endlink[0]{}%
\providecommand \url  [0]{\begingroup\@sanitize@url \@url }%
\providecommand \@url [1]{\endgroup\@href {#1}{\urlprefix }}%
\providecommand \urlprefix  [0]{URL }%
\providecommand \Eprint [0]{\href }%
\providecommand \doibase [0]{http://dx.doi.org/}%
\providecommand \selectlanguage [0]{\@gobble}%
\providecommand \bibinfo  [0]{\@secondoftwo}%
\providecommand \bibfield  [0]{\@secondoftwo}%
\providecommand \translation [1]{[#1]}%
\providecommand \BibitemOpen [0]{}%
\providecommand \bibitemStop [0]{}%
\providecommand \bibitemNoStop [0]{.\EOS\space}%
\providecommand \EOS [0]{\spacefactor3000\relax}%
\providecommand \BibitemShut  [1]{\csname bibitem#1\endcsname}%
\let\auto@bib@innerbib\@empty
\end{thebibliography}%

\end{document}


\title{Supplemental material: Lightly-strained germanium quantum wells with hole mobility exceeding one million}
\author{M. Lodari}
\thanks{These authors contributed equally}
\affiliation{QuTech and Kavli Institute of Nanoscience, Delft University of Technology, PO Box 5046, 2600 GA Delft, The Netherlands}
\author{O. Kong}
\thanks{These authors contributed equally}
\affiliation{School of Physics, University of New South Wales, Sydney, New South Wales 2052, Australia}
\affiliation{ARC Centre of Excellence for Future Low-Energy Electronics Technologies, University of New South Wales, Sydney, New South Wales 2052, Australia}
\author{M. Rendell}
\affiliation{School of Physics, University of New South Wales, Sydney, New South Wales 2052, Australia}
\affiliation{ARC Centre of Excellence for Future Low-Energy Electronics Technologies, University of New South Wales, Sydney, New South Wales 2052, Australia}
\author{A. Tosato}
\affiliation{QuTech and Kavli Institute of Nanoscience, Delft University of Technology, PO Box 5046, 2600 GA Delft, The Netherlands}
\author{A. Sammak}
\affiliation{QuTech and Netherlands Organisation for Applied Scientific Research (TNO), Delft, The Netherlands.}
\author{M. Veldhorst}
\affiliation{QuTech and Kavli Institute of Nanoscience, Delft University of Technology, PO Box 5046, 2600 GA Delft, The Netherlands}
\author{A. R. Hamilton}
\affiliation{School of Physics, University of New South Wales, Sydney, New South Wales 2052, Australia}
\affiliation{ARC Centre of Excellence for Future Low-Energy Electronics Technologies, University of New South Wales, Sydney, New South Wales 2052, Australia}
\author{G. Scappucci*}
\email{g.scappucci@tudelft.nl}
\affiliation{QuTech and Kavli Institute of Nanoscience, Delft University of Technology, PO Box 5046, 2600 GA Delft, The Netherlands}

\date{\today}

\maketitle

\begin{figure}[!ht]
\includegraphics[width=76mm]{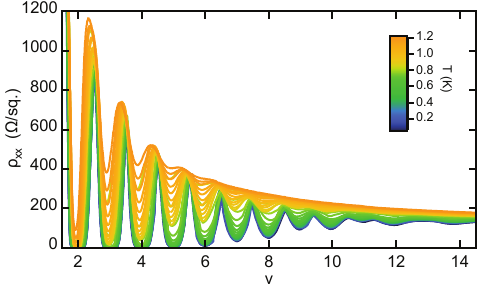}
\caption{\label{fig:1} Temperature dependent longitudinal resistivity $\rho_{xx}$ as a function of filling factor $\nu$ in the temperature range from $T = 68$~mK (blue) to $1.2$~K (orange).}
\end{figure}

We extrapolate the effective $g^*$ factor and the effective mass $m^*$ from the temperature-dependent decay of the longitudinal resistivity $\rho_{xx}$ at different filling factors $\nu = ph/eB_{\nu}$, where $h$ is the Planck constant, $e$ is the electron charge, and $B_{\nu}$ is the magnetic field at integer values of $\nu$. Figure~\ref{fig:1} shows the Shubnikov-de Haas oscillations as a function of $\nu$ measured at a density of $p = 5.8 \times 10^{10}$~cm$^{-2}$ in the temperature range from $T = 68$~mK to $1.2$~K. Filling factors $\nu = 2$ is resolved at a magnetic field $B \sim 1.2$~T, and both even and odd filling factors are resolved up to $\nu = 26$ and $15$ at lower field at the coldest temperature, respectively. 

\begin{figure}[!ht]
\includegraphics[width=76mm]{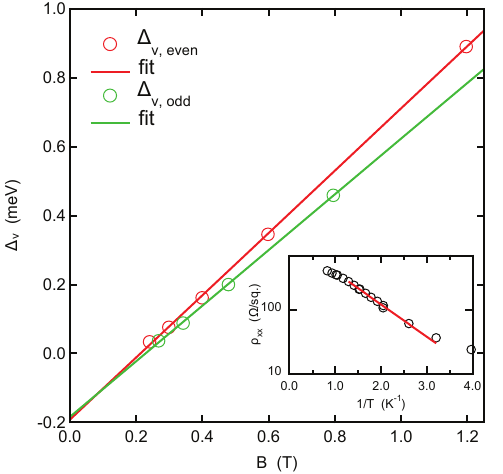}
\caption{\label{fig:2} Activation energy gap $\Delta_{\nu}$ as a function of magnetic field $B$ for even (red) and odd (green) filling factors $\nu$, along with linear fits. The inset shows the Arrhenius plot The inset shows the Arrhenius plot and fit to extract $\Delta_{\nu}$ for $\nu = 5$.}
\end{figure}

The activation energy gap $\Delta_{\nu}$ of each filling factor can be obtained from the thermally activated dependence of the 
SdH oscillation minimum for a given filling factor, as reported in the Arrhenius plot of the $ln(\rho_{xx,\nu})$ vs. $1/T$ (black circles, inset Fig.~\ref{fig:2}). Following the Boltzmann statistics, the longitudinal magnetoresistance of a specific minima can be described via the relation $ln(\rho_{xx,\nu}) \propto - \Delta_{\nu}/(2k_BT)$, where $k_B$ is the Boltzmann constant. The activation energy of the filling factors $\Delta_{\nu}$ can be extrapolated from the slope of a linear fit (red).\\
Since the even and odd filling factors corresponds to the cyclotron frequency and the Zeeman splitting, respectively, and a linear relation links activation energy $\Delta_{\nu}$ and the magnetic field $B_{\nu}$ at which each $\nu$ occur, the effective mass $m^*$ and the effective $g^*$ can be extrapolated from a linear fit of the $\Delta_{\nu} (B_{\nu})$ dependence.
Figure~\ref{fig:2} shows the extrapolated activation energy $\Delta_{\nu}$ as a function of magnetic field $B$ for all the investigated even and odd filling factors (red and green, respectively). The effective $g^*$ can be extrapolated from the slope of a linear fit $\Delta_{\nu,odd} = g^*\mu_B B$ (green line), where $\mu_B$ is the Bohr magnetron. Once the $g^*$ is know, then the effective mass $m^*$ can be obtained from the slope of the linear fit $\Delta_{\nu,even} = \hbar eB/m^* - g^*\mu_B B$ (red line).\\
For a density of $p = 0.58\times10^{10}$~cm$^{-2}$, we found an effective $g^*$ factor of $13.95\pm0.18$ and an effective mass $m^*$ of $0.068\pm0.001$.\\

